\documentclass[ twocolumn, 
                secnumarabic, 
                amssymb, 
                nobibnotes,
                superscriptaddress,
                aps, 
                prl, 
                shortbibliography]{revtex4-2}
\usepackage{color, 
            soul, 
            graphicx, 
            amsmath,
            amsbsy,
            etoolbox,
            natbib,
            hyperref}

\usepackage{pdfpages}

\makeatletter
\AtBeginDocument{\let\LS@rot\@undefined}
\makeatother

\begin{document}

\title{Quantum Lifetime Spectroscopy and Magnetotunneling in Double Bilayer Graphene Heterostructures}%

\author{Nitin~Prasad}
\altaffiliation[Present Address: ]{Department of Chemistry and Biochemistry, University of Maryland, College Park, MD 20742, USA}
\affiliation{Microelectronics Research Center, Department of Electrical and Computer Engineering, The University of Texas at Austin, Austin, TX 78758, USA}
\author{G.~William~Burg}
\affiliation{Microelectronics Research Center, Department of Electrical and Computer Engineering, The University of Texas at Austin, Austin, TX 78758, USA}
\author{Kenji~Watanabe}
\affiliation{Research Center for Functional Materials, National Institute of Materials Science, 1-1 Namiki Tsukuba, Ibaraki 305-0044, Japan}
\author{Takashi~Taniguchi}
\affiliation{International Center for Materials Nanoarchitectonics, National Institute of Materials Science, 1-1 Namiki Tsukuba, Ibaraki 305-0044, Japan}
\author{Leonard~F.~Register}
\affiliation{Microelectronics Research Center, Department of Electrical and Computer Engineering, The University of Texas at Austin, Austin, TX 78758, USA}
\author{Emanuel~Tutuc}
\email[Corresponding Author: ]{etutuc@mer.utexas.edu}
\affiliation{Microelectronics Research Center, Department of Electrical and Computer Engineering, The University of Texas at Austin, Austin, TX 78758, USA}

\date{\today}%

\begin{abstract}
We describe a tunneling spectroscopy technique in a double bilayer graphene heterostructure where momentum-conserving tunneling between different energy bands serves as an energy filter for the tunneling carriers, and allows a measurement of the quasi-particle state broadening at well defined energies. The broadening increases linearly with the excited state energy with respect to the Fermi level, and is weakly dependent on temperature. In-plane magnetotunneling reveals a high degree of rotational alignment between the graphene bilayers, and an absence of momentum randomizing processes.
\end{abstract}

\maketitle

Energy resolved lifetime measurements provide unique insight into the fundamental relaxation mechanisms of quantum states in a material. The carrier lifetime, the inverse of quasi-particle state broadening is however elusive to transport, which probes primarily momentum relaxation. Angle-resolved photoemission spectroscopy \cite{Xu_ARPES_1996, Perfetti_ARPES_2001, Haberer_anisotropic_2013} has been generally used to been used to extract the state carrier lifetime, and in certain conditions scanning tunneling spectroscopy can probe the carrier lifetimes for discrete states, such as Landau levels in graphene \cite{Li_STM_2009}.  
The tunneling spectra between two closely spaced two-dimensional (2D) layers offers a unique tool to probe lifetimes of carriers with continuously varying energies above and below the Fermi surface. Quantum tunneling is an energy conserving phenomenon, and in translation invariant systems where momentum is a good quantum number in both layers, momentum-conserving tunneling leads to resonances in the interlayer tunneling conductance \cite{eisenstein_resonant_tunneling_1991, feenstra_single_particle_2012, kang_itfet_review_2017}. Studies in in GaAs double quantum wells of electrons \cite{turner_tunneling_2D_1993, murphy_lifetime_2D_1995} and holes \cite{hayden_magnetotunneling_gaas_1991, eisenstein_lifetime_2D_2007} shed light on Fermi surface properties and carrier lifetime.

In this letter, we use momentum-conserving tunneling between two rotationally aligned graphene bilayers, henceforth referred to as \textit{bilayers}, to investigate quasi-particle state broadening in the individual bilayers, as a function of energy and temperature.  We use magnetotunneling spectroscopy to confirm the high degree rotational alignment of the two bilayers, and determine the bilayer spatial separation. The data reveal the quasi-particle state broadening increases linearly with the state energy separation from the Fermi level, and is relatively insensitive to temperature, which in turn suggests the quasi-particle state broadening at the Fermi level is disorder limited, and at excited states is dominated by carrier-carrier interactions or phonon emission.

\begin{figure}
    \centering
    \includegraphics[width=1.0\linewidth]{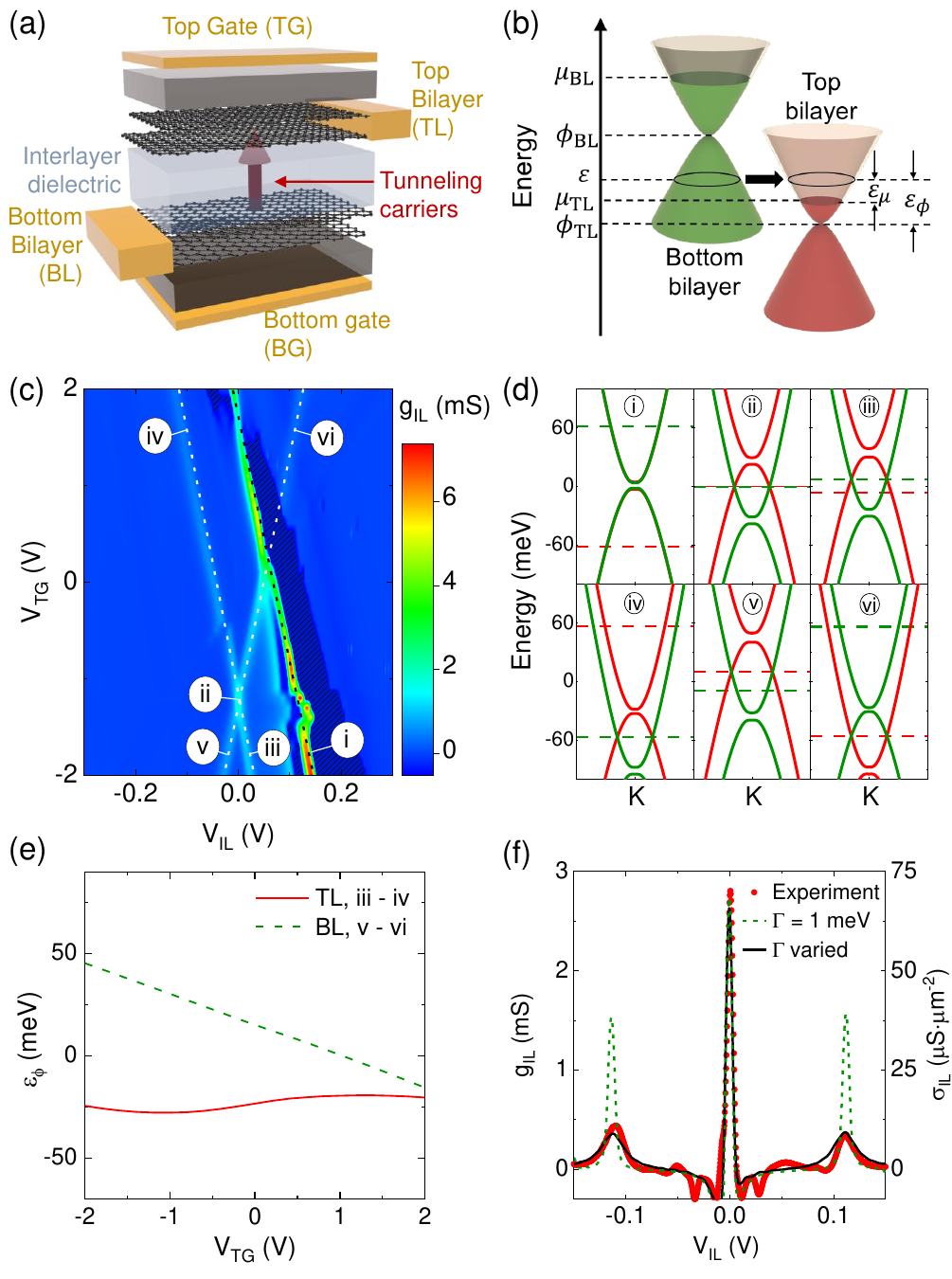}
    \caption{(a) Double bilayer graphene heterostructure schematic. (b) Top and bottom bilayer band alignment for a general biasing condition. (c) $g_\mathrm{IL}$ vs. $V_\mathrm{TG}$ and $V_\mathrm{IL}$ at $T$~=~1.5~K for Device~\#1. The shaded region is experimentally inaccessible due to the negative differential resistance-induced circuit instability \cite{burg_coherent_interlayer_2017}. The six feature points (i-vi) in panel (c) correspond to the six different tunneling regimes. (d) 
    Self-consistently calculated top (red) and bottom (green) bilayer bandstructures around the \textbf{K}-points corresponding to points (i-vi) in panel (c). The dashed lines are the chemical potentials of the respective layers.
    (e) $\varepsilon_\phi$ vs. $V_\mathrm{TG}$ along the contours (iii)-(iv) and (v)-(vi) indicated by white dotted lines in (c). (f) $g_\mathrm{IL}$ vs. $V_\mathrm{IL}$ (red dots) measured in Device \#1 at $T = 1.5$~K. The primary resonance occurs at $V_\mathrm{IL} = 0$~V, and secondary resonances are visible at $V_\mathrm{IL} \approx \pm 0.12$~V. The green-dotted (black-solid) line represents calculations with a constant (variable) broadening. The right axis shows $g_\mathrm{IL}$ normalized to the overlap area ($\sigma_\mathrm{IL}$).}
    \label{fig:Figure1}
\end{figure}

The two bilayers are defined from a large-area single crystal using electron-beam lithography (EBL) and $\mathrm{O_2}$ plasma etching, which ensures crystallographic alignment from the outset of fabrication. As shown in Fig. \ref{fig:Figure1}(a), a heterostructure consisting of double bilayers separated by a few-layer thick WSe\textsubscript{2} tunnel barrier is created by using a polymer stamp to pick up and stack individual layers, while maintaining the original rotational orientation of the bilayers \cite{kim_van_2016}. Multiple independent contacts to each bilayer, and metal top and bottom gates are defined using EBL and metal evaporation. The dual-gate geometry allows for independent control of the charge densities in each bilayer. We consider here two heterostructures, one with a two-layer $\mathrm{WSe_2}$ (Device \#1) and one with a three-layer $\mathrm{WSe_2}$ (Device \#2) tunnel barrier.

As we will show in this study, using a combination of measurements and calculations, the carriers tunneling between the graphene bilayers conserve both energy and momentum thanks to the accurate rotational alignment of the two bilayers. To gain an insight into the measured interlayer tunneling currents ($I_\mathrm{IL}$), we use a single-particle tunneling model \cite{zheng_tunneling_2D_1993, turner_tunneling_2D_1993, burg_coherent_interlayer_2017}.
\begin{equation}
    I_\mathrm{IL} = -e \int_{-\infty}^{\infty}\mathrm{d}E\,\, T(E)(f(E-\mu_\mathrm{TL}) - f(E-\mu_\mathrm{BL})) \label{eq:current}
\end{equation}
where $f(E)$ is the Fermi distribution function, $e$ the elementary charge, $\mu_\mathrm{TL}$ ($\mu_\mathrm{BL}$) the chemical potential of the top (bottom) bilayer, and $T(E)$ the vertical transmission rate at energy $E$ . Assuming a weak interlayer coupling, $T(E)$ is given by
\begin{equation}
    T(E) = \frac{2\pi}{\hbar}\sum\limits_{\substack{\mathbf{k}_\mathrm{TL}, \mathbf{k}_\mathrm{BL};\\s_\mathrm{TL},s_\mathrm{BL}}}|t|^2 A_{\mathbf{k}_\mathrm{TL},s_\mathrm{TL}}(E)A_{\mathbf{k}_\mathrm{BL},s_\mathrm{BL}}(E) \delta_{\mathbf{k}_\mathrm{TL} \mathbf{k}_\mathrm{BL}}. \label{eq:transmission}
\end{equation}
The wavevectors $\mathbf{k}_\mathrm{TL}$ ($\mathbf{k}_\mathrm{BL}$) span the first Brillouin zone, and $s_\mathrm{TL}$ ($s_\mathrm{BL}$) the first two valence and conduction subbands  of the top (bottom) bilayer. The interlayer coupling parameter $t$ is assumed independent of $\mathbf{k}$ and $E$, and depends only on the interlayer separation. The spectral density functions of the top ($A_\mathrm{TL}$) and bottom ($A_\mathrm{BL}$) bilayers are given by
\begin{equation}
    A_{\mathbf{k},s}(E) = \frac{1}{\pi}\left(\frac{\Gamma}{\left(E-\varepsilon_{\mathbf{k},s}\right)^2 + \Gamma^2}\right) \label{eq:broadening}
\end{equation}
where $\Gamma$ is energy broadening of the quasi-particle state.  We compute the energy dispersion $\varepsilon_{\mathbf{k},s}$ within each bilayer using a $p_z$ orbital-based tight-binding model \cite{mccann_blg_tightbinding_2013}, with  the parameters of Ref. \cite{kuzmenko_experimental_tightbinding_2009}, without including the trigonal warping. We self-consistently model the band-gaps within the bilayers in the presence of transverse electric fields \cite{min_abinitio_blg_2007}. A schematic bandstructure of the top and bottom bilayers at a given interlayer bias voltage $V_\mathrm{IL} = -(\mu_\mathrm{TL} - \mu_\mathrm{BL})/e$, and top (bottom) gate voltage $V_\mathrm{TG}$ ($V_\mathrm{BG}$) is shown in Fig. \ref{fig:Figure1}(b). The electrostatic potentials of the top ($\phi_\mathrm{TL}$) and bottom ($\phi_\mathrm{BL}$) bilayers, which control the relative alignment of the energy bands are calculated using a capacitive model \cite{burg_coherent_interlayer_2017} (Table \ref{tab:capacitances}).

\begin{table}[h]
\caption{\label{tab:capacitances}Top gate ($C_\mathrm{TG}$), bottom gate ($C_\mathrm{BG}$), and interlayer ($C_\mathrm{IL}$) capacitances for Devices \#1 and \#2.}
\begin{ruledtabular}
\begin{tabular}{lccc} 
    & $C_\mathrm{TG}$ ($\mathrm{\mu F/cm^2}$) & $C_\mathrm{IL}$ ($\mathrm{\mu F/cm^2}$) & $C_\mathrm{BG}$ ($\mathrm{nF/cm^2}$)     \\ 
\hline
Device \#1    & $0.19$ & $2.0$ & $11$  \\
Device \#2    & $0.16$ & $1.3$ & $72$  \\
\end{tabular}
\end{ruledtabular}
\end{table}

Figure \ref{fig:Figure1} (c) shows an example of the measured interlayer tunneling conductance $g_\mathrm{IL}=\partial I_\mathrm{IL}/\partial V_\mathrm{IL}$ as a function of $V_\mathrm{IL}$ and $V_\mathrm{TG}$, at $V_\mathrm{BG}=25$~V and a temperature $T=1.5$~K. The data reveal a set of resonances, marked by $g_\mathrm{IL}$ maxima, which form nearly linear contours in the $V_\mathrm{IL}$-$V_\mathrm{TG}$ plane. The bilayers' bands and chemical potentials at six representative points labelled (i)-(vi) are shown in Fig. \ref{fig:Figure1} (d).
The black-dotted line in Fig. \ref{fig:Figure1} (c) corresponds to the primary resonance, where the bands of the two bilayers align [Fig. \ref{fig:Figure1} (d)(i)]. For a general biasing condition the bands of the top and bottom bilayers are misaligned, and momentum-conserving tunneling is suppressed except at particular a energy ($\varepsilon$) corresponding to a ring of states formed by the intersection of the electron band of one bilayer and hole band of the other, as depicted in Fig.~\ref{fig:Figure1}~(b). We refer to this regime as unlike-band tunneling, as opposed to the like-band tunneling which controls the primary resonance in Fig.~\ref{fig:Figure1}~(c). Unlike-band tunneling is possible only if $\varepsilon$ lies between $\mu_\mathrm{TL}$ and $\mu_\mathrm{BL}$. Along the two white-dotted lines in Fig. \ref{fig:Figure1} (c), the ring of intersection at $\varepsilon$ crosses into the interval between $\mu_\mathrm{TL}$ and $\mu_\mathrm{BL}$ [Fig.~\ref{fig:Figure1}~(d) (ii)-(vi)], leading to secondary resonances in the tunneling conductance. Particularly, $\varepsilon=\mu_\mathrm{BL}$ along the (iii)-(iv) line, and $\varepsilon=\mu_\mathrm{TL}$ along the (v)-(vi) line. Figure \ref{fig:Figure1} (e) shows $\varepsilon_\phi$, namely the energy of the ring of intersection referenced to the charge neutrality of the bilayer whose chemical potential is not aligned with $\varepsilon$ [Fig. \ref{fig:Figure1}(b)], as a function of $V_\mathrm{TG}$ along the lines (iii)-(iv) and (v)-(vi). The $\varepsilon_\phi$ values along (iii)-(iv) are almost constant, indicating that along the (iii)-(iv) contour carriers are injected to the same set of states in the top bilayer.

Figure \ref{fig:Figure1}(f) shows the measured $g_\mathrm{IL}$ vs. $V_\mathrm{IL}$, at $V_\mathrm{TG}~=~1.7$~V and $V_\mathrm{BG}~=~25$~V, adjusted so that the carrier densities in the top ($n_\mathrm{TL}$), and bottom ($n_\mathrm{BL}$) bilayers are $n_\mathrm{TL} = n_\mathrm{BL} = 1.7 \times 10^{12}\,\mathrm{cm^{-2}}$ at $V_\mathrm{IL} = 0$~V, and the primary resonance condition occurs at $V_\mathrm{IL} = 0$~V. A fit using Eqs. (\ref{eq:current}-\ref{eq:broadening}) around the primary resonance yields $\Gamma=1$~meV and $t=30$~$\mu$eV. 
However, the model of Eqs. (\ref{eq:current}-\ref{eq:broadening}) using a fixed $\Gamma=1$~meV does not accurately capture the secondary resonances at $V_\mathrm{IL}\approx \pm 0.12$~V, corresponding to points (iv) and (v) in Fig. \ref{fig:Figure1}(c). Indeed, the widths (peak values) of the secondary resonances are underestimated (overestimated), indicating that the quasi-particle state broadening varies depending on the energy or relative position to the chemical potential. Additional data supporting this conclusion are presented in Fig. S1 of the Supplementary Material.

Quasi-particle states can be broadened by relaxation mechanisms, such as static disorder, phonon emission, and carrier-carrier interactions. Their relative contributions depend on the state energy, band structure, chemical potential, and temperature. Ignoring trigonal warping,     
$\Gamma$ reduces to a function of the energy separation from the Fermi surface ($\varepsilon_\mu = \varepsilon - \mu$) and the energy referenced to the charge neutrality level ($\varepsilon_\phi = \varepsilon - \phi$) [Fig. \ref{fig:Figure1}(b)], $\Gamma \equiv \Gamma(\varepsilon_\mu, \varepsilon_\phi)$. When the primary resonance occurs at $V_\mathrm{IL}=0$~V, the tunneling is restricted to a narrow range of energies at the Fermi surfaces of the two bilayers, and the width of the primary resonance can be used to extract $\Gamma = \Gamma(0, \varepsilon_\phi)$. 

The secondary resonances are particularly interesting, as momentum-conserving unlike band tunneling creates a filter for carrier injection at a single energy $\varepsilon$. Along the (iii)-(iv) contour in Fig. \ref{fig:Figure1}(c), $\varepsilon = \mu_\mathrm{BL}$, and the contributing values of broadening to transmission rate in \eqref{eq:transmission} correspond to $\Gamma(\mu_\mathrm{BL} - \mu_\mathrm{TL}, \mu_\mathrm{BL}-\phi_\mathrm{TL})$ of the top bilayer and $\Gamma(0, \mu_\mathrm{BL}-\phi_\mathrm{BL})$ of the bottom bilayer. Therefore, by performing a fit to the experimental data at the secondary resonances we can extract the broadening of quantum states away from the Fermi surface. In this case, the bottom bilayer is either injecting particles into empty states [point (iii)], or vacancies into filled states of the top bilayer [point (iv)]. Similarly,  along the (v)-(vi) contour in Fig. \ref{fig:Figure1}(c), $\varepsilon = \mu_\mathrm{TL}$, which allows the extraction of the bottom bilayer quantum state broadening. The model fits  in Fig. \ref{fig:Figure1}(f) with $\Gamma=1$~meV and $\Gamma=12$~meV used for the primary and secondary resonances, respectively, are in excellent agreement with the experimental data.

\begin{figure}
    \centering
    \includegraphics[width=\linewidth]{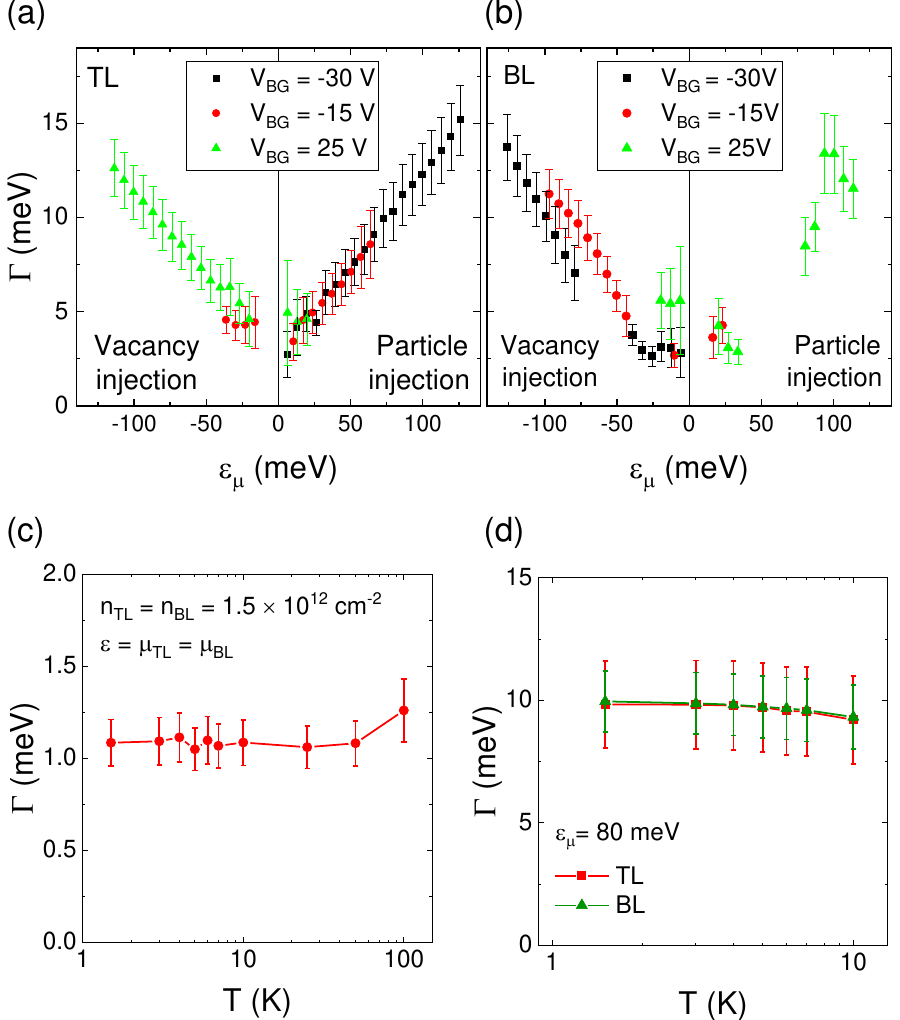}
    \caption{(a,b) $\Gamma$ vs. $\varepsilon_\mu$ in the top [panel (a)] and bottom [panel (b)] bilayer for three $V_\mathrm{BG}$ values. (c) $\Gamma$ vs. $T$ at $\varepsilon_\mu=0$ and $\varepsilon_\phi=60$~meV. (d) $\Gamma(80\mathrm{\,meV}, 23\mathrm{\,meV})$ vs. $T$ of the top and bottom bilayers extracted at the secondary resonances. The $\Gamma$ values in panels (c,d) are relatively insensitive to temperature. The error bars indicate the standard fitting error.}
    \label{fig:Figure2}
\end{figure}

Figure \ref{fig:Figure2}(a) shows the top bilayer $\Gamma$ vs. $\varepsilon_\mu$, extracted from parameter fits along (iii)-(iv) contour, at various $V_\mathrm{BG}$. As established in Fig. \ref{fig:Figure1}(e), $\varepsilon_\phi$ is constant along (iii)-(iv) at a fixed $V_\mathrm{BG}$. Therefore, the back-gate bias provides an independent handle to vary the $\varepsilon_\phi$, and $V_\mathrm{BG}=-30$~V, $-15$~V, and $25$~V correspond to $\varepsilon_\phi \approx 35$~meV, $17$~meV and $-24$~meV respectively. Remarkably, Fig. 2(a) data show that $\Gamma$ does not vary significantly with $\varepsilon_\phi$. For $\varepsilon > \mu_\mathrm{TL}$ ($\varepsilon < \mu_\mathrm{TL}$), the bottom bilayer injects particles (vacancies) into the top bilayer, and the extracted $\Gamma$ corresponds to the particle (vacancy) state broadening. Similarly, Fig. \ref{fig:Figure2}(b) shows the bottom bilayer $\Gamma$ vs. $\varepsilon_\mu$, from parameter fits along the (v)-(vi) contour, at various $V_\mathrm{BG}$. We note that unlike the (iii)-(iv) contour, the  $\varepsilon_\phi$ varies along the (v)-(vi) contour [Fig. \ref{fig:Figure1}(e)], concomitantly with $\varepsilon_\mu$. Nonetheless, Fig. \ref{fig:Figure2}(a) and (b) data are in good agreement, and show that the $\Gamma$ values are largely controlled by $\varepsilon_\mu$, and scale almost linearly with $\varepsilon_\mu$.

\begin{figure*}
    \centering
    \includegraphics[width=1.0\linewidth]{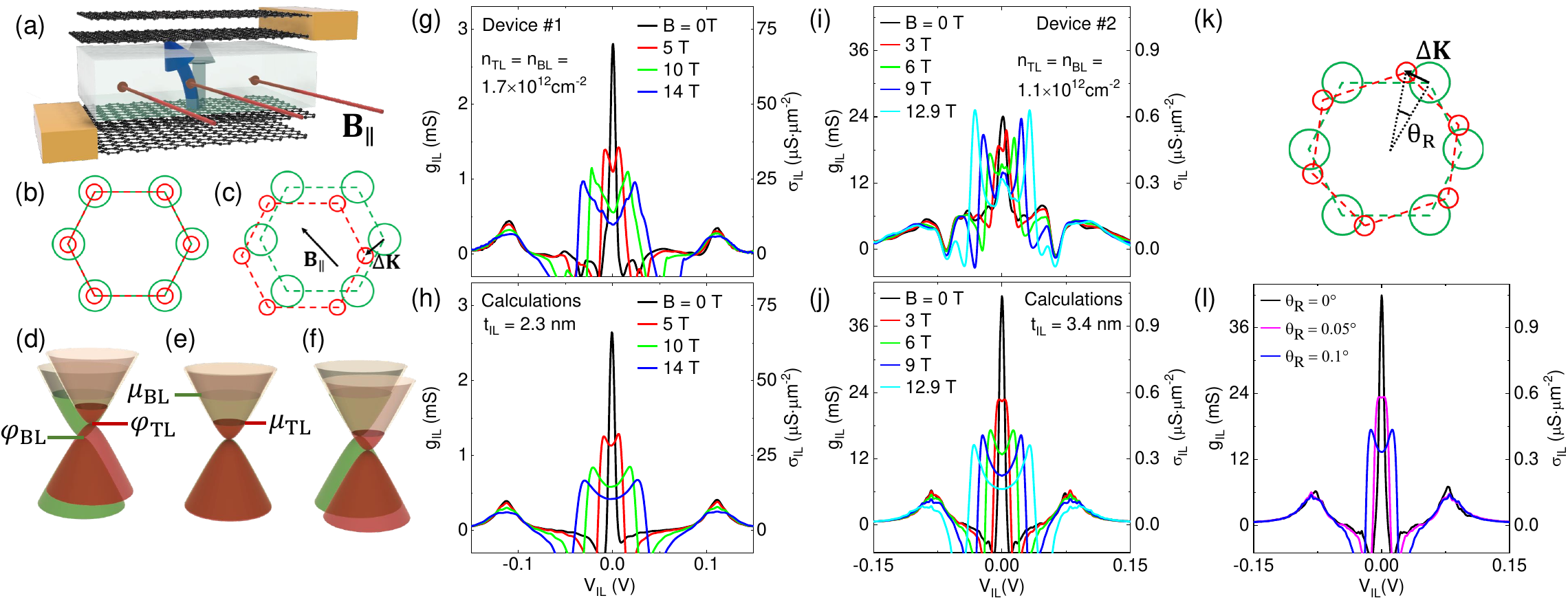}
    \caption{(a) Schematic of the tunneling process in the presence of an in-plane magnetic field, where carriers gain an additional momentum through the Lorentz force. (b-c) Fermi surfaces (solid lines) and first Brillouin zones (dashed lines) of the top (red) and bottom (green) bilayers (b) without  and (c) with an applied $\mathbf{B}_\mathrm{\parallel}$. (d-f) Band-alignment around the $\mathbf{K}$-points in the presence of $\mathbf{B}_\mathrm{\parallel}$. At $\mathbf{B}_\mathrm{\parallel}=0$ [panel (e)] the band-alignment corresponds to a primary resonance. An applied $\mathbf{B}_\mathrm{\parallel}$ leads to a splitting of the resonance shown in panels (d) and (f). (g-j) Experimental [panels (g),(i)] and calculated [panels (h),(j)] $g_\mathrm{IL}$ (left axis) and $\sigma_\mathrm{IL}$ (right axis)  vs. $V_\mathrm{IL}$ in Device \#1 [panels (g-h)] and Device \#2 [panels (i-j)] at various $B$, and at $T = 1.5$~K. The layer densities at $V_\mathrm{IL} = 0$~V are indicated in (g) and (i). The data show a clear of the resonance splitting, which is used to determine $t_\mathrm{IL}$. (k) Brillouin zones of the top (red) and bottom (green) bilayers in the presence of a small relative twist $\theta_\mathrm{R}$. (l) Calculated $g_\mathrm{IL}$ (left axis) and $\sigma_\mathrm{IL}$ (right axis)  vs. $V_\mathrm{IL}$ for Device \#2 at $B$~=~0 T for different $\theta_\mathrm{R}$ values.}
    \label{fig:Figure3}
\end{figure*}

To understand the mechanism controlling the quasi-particle state broadening, we also consider the $\Gamma$ vs. $T$ dependence shown in Fig. \ref{fig:Figure2}(c,d). The primary resonance at the Fermi surface ($\varepsilon_\mu=0$) with $n_\mathrm{TL} = n_\mathrm{BL} = 1.5\times10^{12}\mathrm{\,cm}^{-2}$ ($\varepsilon_\phi = 60$~meV) exhibits a weak temperature dependence up to $T=100$ K [Fig. \ref{fig:Figure2}(c)], which indicates that it is controlled primarily by disorder, as acoustic phonon scattering leads to a $\Gamma \propto T$ or stronger dependence \cite{Hwang_scattering_2008, Ma_scattering_2014}, and carrier-carrier interaction to a $\Gamma \propto T^2$ dependence \cite{murphy_lifetime_2D_1995, eisenstein_lifetime_2D_2007}. Figure \ref{fig:Figure2}(d) shows $\Gamma$ vs. $T$ away from the Fermi surface, determined from the secondary resonances at $\varepsilon_\mu = 80$~meV. In both Fig. \ref{fig:Figure2}(c,d) data $\Gamma$ is nearly insensitive to $T$. Overall Fig. 2 data indicate that $\Gamma$ at the Fermi level is controlled by disorder in a wide temperature range, and away from the Fermi level by carrier-carrier interaction and phonon emission \cite{Park_Linewidth_2009, park_inelastic_2012}. We note that the relative contributions of carrier-carrier interaction and phonon emission have yet to be clarified theoretically, and Fig. 2 data provide an experimental point of comparison for such study. 

In light of finite $I_\mathrm{IL}$ and $V_\mathrm{IL}$ values used in the tunneling measurements it is instructive to examine the role of electron heating. Using a maximum dissipated power of 5 W/cm$^2$ at the secondary resonances in our experiments, we estimate the lattice temperature increase to be less than 0.1~K \cite{choi_heating_2018, seol_phonon_2010}. Following calculations similar to Ref. \cite{vallabhaneni_phonon_2016} we find that the difference between the electron and lattice temperatures, controlled by the electron-phonon coupling, to be less than 0.5~K in the range of biases used to extract the $\Gamma$ values. Since $\Gamma$ is insensitive to temperatures up to 10 K, we conclude that Joule heating does not significantly affect the reported results.

A key ingredient in our analysis has been the assumption of momentum-conserving tunneling enabled by the rotational alignment of the two bilayers. To validate this hypothesis, we examine the tunneling characteristics with an applied in-plane magnetic field ($\mathbf{B}_\parallel$) [Fig. \ref{fig:Figure3}(a)], which adds a momentum to the tunneling carrier \cite{mishchenko_twist_controlled_2014, wallbank_bilayer_2016} [Fig. \ref{fig:Figure3}(b-c)]:  \begin{equation}
    \hbar\Delta \mathbf{K} = e\,t_\mathrm{IL} (\hat{\textbf{z}} \times \mathbf{B}_\mathrm{\parallel}), \label{eq:kshift}
\end{equation}
and leads to a splitting of the primary resonance [Fig. \ref{fig:Figure3}(d-f)]; $t_\mathrm{IL}$ is the tunneling distance, and $\hat{\textbf{z}}$ the unit vector perpendicular to the sample plane. 

Figures \ref{fig:Figure3}(g) and \ref{fig:Figure3}(i) show the measured $g_\mathrm{IL}$ vs. $V_\mathrm{IL}$  for Devices \#1 and \#2, respectively,  at various $B = |\textbf{B}_\mathrm\parallel|$, $T = 1.5$~K, and at biasing conditions 
such that the primary resonance is at $V_\mathrm{IL}=0$~V. In both panels, the data show a clear splitting of the resonance with increasing $B$, with the two peaks occur at the band alignment conditions of Figs.~\ref{fig:Figure3}(d) and ~\ref{fig:Figure3}(f). Figures \ref{fig:Figure3}(h) and \ref{fig:Figure3}(j) show the calculated $g_\mathrm{IL}$ for the same set of $V_\mathrm{IL}$ and $B$ as in Figs. \ref{fig:Figure3}(g) and \ref{fig:Figure3}(i), respectively. Figure S2 of the Supplementary Material expands the comparison of the experimental and calculated $g_\mathrm{IL}$ vs. $V_\mathrm{IL}$ and $V_\mathrm{TG}$ in Device \#1. The applied $\mathbf{B}_\mathrm{\parallel}$ enters Eq. \eqref{eq:broadening} as a wave-vector shift between the top and bottom bilayers $\mathbf{k}_\mathrm{TL} - \mathbf{k}_\mathrm{BL} = \Delta\mathbf{K}$. While we account for the bilayers energy bandstructure change in the presence of $\mathbf{B}_\mathrm{\parallel}$ according to Ref. \cite{pershoguba_parallel_graphene_2010}, no appreciable changes in the model results are observed in the $B$-range experimentally explored. By matching the calculated resonance splitting as a function of $B$ we extract a tunneling distance $t_\mathrm{IL}=2.3$ nm and $t_\mathrm{IL}=3.4$ nm for Devices \#1 and \#2, respectively. These values are in good agreement with the tunnel barriers of two and three WSe$_2$ layers for Devices \#1 and \#2, respectively, albeit larger than the expected tunnel barrier thickness. We note that including the trigonal warping in the bilayer graphene bandstructure does not change the $g_\mathrm{IL}$  values at $B=0$ T, but does lead to additional features in the magnetotunneling characteristics (Fig. S3 of the Supplementary Material).

Lastly, we address the impact of a small twist ($\theta_\mathrm{R}$) between the top and bottom bilayers, which has a similar effect as applied $\mathbf{B}_\mathrm{\parallel}$ for small angles \cite{he_twist_pseudomagnetic_2014}, and changes Eq. \eqref{eq:kshift} to:
    $\hbar\Delta \mathbf{K} =  \hat{\textbf{z}} \times(e\,t_\mathrm{IL} \mathbf{B}_\mathrm{\parallel} + \theta_\mathrm{R}\hbar\mathbf{K}) \label{eq:kshiftandrotate}$.
The momentum shift is valley dependent in this case. Figure~\ref{fig:Figure3}(l) shows the calculated $g_\mathrm{IL}$ vs. $V_\mathrm{IL}$ for different $\theta_\mathrm{R}$, corresponding to the $B = 0$~T case of Fig.~\ref{fig:Figure3}(j). We note here that even a small, $\theta_\mathrm{R}=0.1^\circ$ twist between the top and bottom layers is expected to introduce a significant splitting of the resonance $g_\mathrm{IL}$ peak. Because no splitting is observed at $B = 0$~T in the experimental data of Fig.~\ref{fig:Figure3}(g) and Fig.~\ref{fig:Figure3}(i), we conclude that both Device \#1 and \#2 have a high degree of rotational alignment between the bilayers. 

In summary, we describe a spectroscopy technique where momentum-conserving tunneling between different bands in a double layer heterostructure acts as an energy filter, and allows the extraction of quasi-particle state broadening at a well defined energy with respect to the Fermi level. The technique leverages advances in van der Waals heterostructures fabrication with exquisite control of rotational alignment, as demonstrated by in-plane magnetotunneling.  

\begin{acknowledgments}
We thank A. H. MacDonald, F. Giustino, and Li Shi for useful discussions. The work at The University of Texas was supported by the National Science Foundation Grants No. EECS-1610008 and NSF-MRSEC DMR-1720595, Army Research Office under Grant No. W911NF-17-1-0312, and the Welch Foundation grant F-2018-20190330. Work was partly done at the Texas Nanofabrication Facility supported by NSF Grant No. NNCI-1542159, and at the Texas Advanced Computing Center (TACC) at The University of Texas at Austin. K.W. and T.T. acknowledge support from the Elemental Strategy Initiative conducted by the MEXT, Japan, Grant Number JPMXP0112101001,  JSPS KAKENHI Grant Number  JP20H00354 and the CREST(JPMJCR15F3), JST. 
\end{acknowledgments}

\bibliographystyle{apsrev4-2}

%

\newpage
\includepdf[pages={1}]{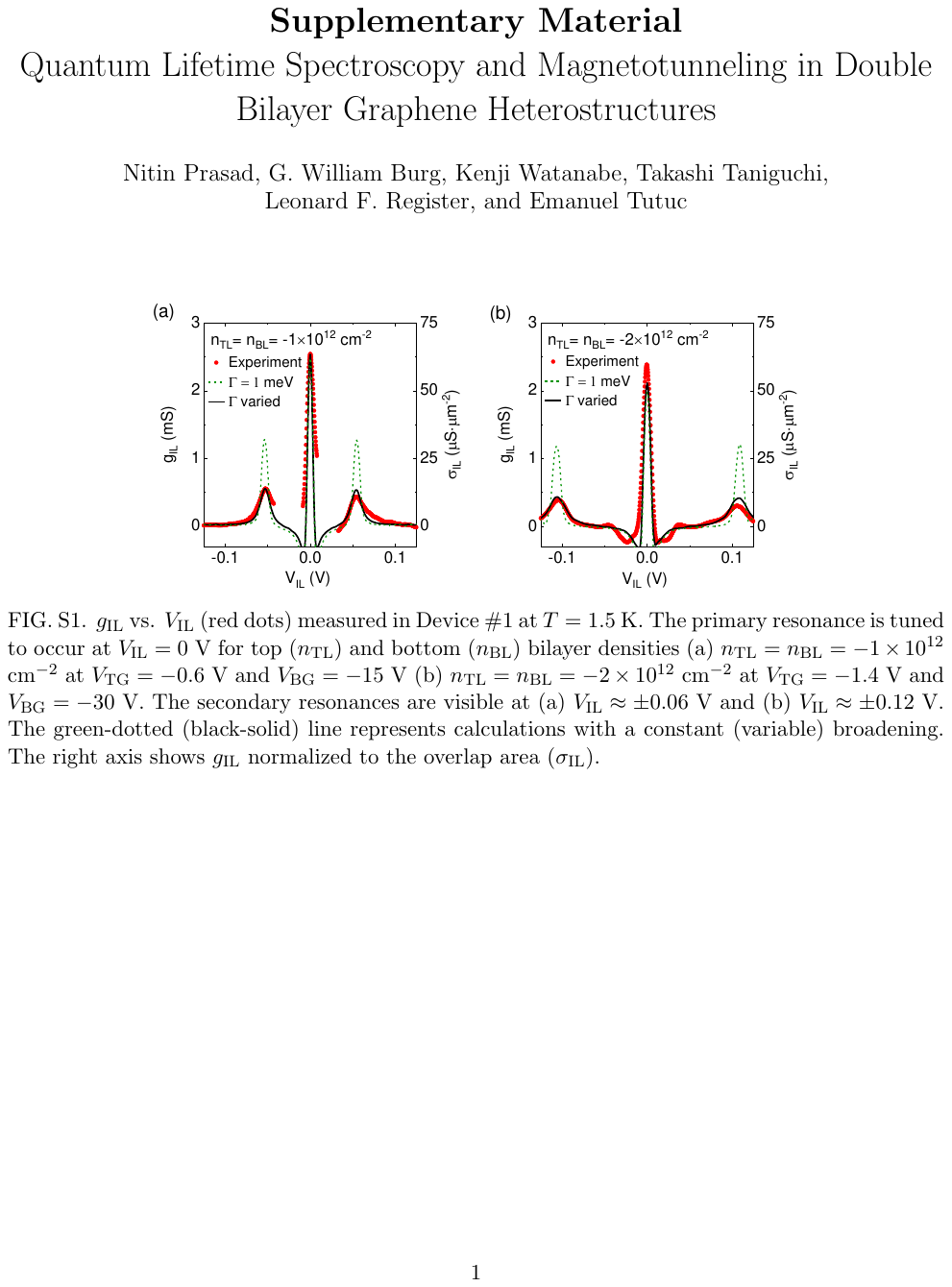}
{\color{white} .}
\newpage
\includepdf[pages={2}]{SM.pdf}
{\color{white} .}
\newpage
\includepdf[pages={3}]{SM.pdf}

\end{document}